# 3D Single-shot CEST imaging at 3T Based on True FISP Readout

Yupeng Wu [a], Qifan Pang [a], Zhichao Wang [b], Gaiying Li [a], Caixia Fu [c], Mengqiu Cao [d], Xingrui Wang [d], Dongmei Jiang [e], Dejun She [e], Yang Song [f], Yu Zhao [g,h,i]*, Jianqi Li [a]*

[a] Shanghai Key Laboratory of Magnetic Resonance, School of Physics and Electronic Science, East China Normal University, Shanghai, China

[b] Zhejiang Lab, Hangzhou, Zhejiang, China

[c] MR Collaboration, Siemens (Shenzhen) Magnetic Resonance, Shenzhen, China

[d] Department of Radiology, Renji Hospital, School of Medicine, Shanghai Jiao Tong University, Shanghai, China

[e] Department of Radiology, First Affiliated Hospital of Fujian Medical University, Fuzhou, China

[f] MR Scientific Marketing, Siemens Healthineers, Shanghai, China

[g] Department of Radiology, and Functional and Molecular Imaging key Laboratory of Sichuan Province, West China Hospital of Sichuan University, Chengdu, China

[h] Huaxi MR Research Center (HMRRC), West China Hospital of Sichuan University, Chengdu, China

[i] Research Unit of Psychoradiology, Chinese Academy of Medical Sciences, Chengdu, China

***Corresponding authors:**

Jianqi Li, Ph.D, Shanghai Key Laboratory of Magnetic Resonance, School of Physics and Electronic Science, East China Normal University, 3663 North Zhongshan Road, Shanghai, 200062, China; Email: jqli@phy.ecnu.edu.cn.

Yu Zhao, Ph.D, Department of Radiology, and Functional and Molecular Imaging key Laboratory of Sichuan Province, West China Hospital of Sichuan University, No.37 Guoxue Lane, Chengdu, 610041, China; Email: zhaoyu2013022063@163.com.

**Word count:** 5764

**Abstract**

To simultaneously fit multiple-pool effects, spectrally selective 3D chemical exchange saturation transfer (CEST) imaging typically requires single-shot readouts to save time. However, to date, fast low angle shot (FLASH) and echo planar imaging (EPI) have been the primary pulse sequences used for this purpose. They suffer from low signal-to-noise ratio (SNR) or image distortion related to B0 field inhomogeneity. In this work, we developed a 3D single-shot CEST sequence using true fast imaging with steady-state precession (True FISP) readout, also known as balanced steady state free precession (bSSFP), and optimized the scanning parameters through simulations. The performance of the CEST sequence was validated using an egg white phantom, ten healthy volunteers, and a patient with a brain tumor on a 3T human scanner. Subsequently, the proposed CEST sequence using True FISP was compared with the commonly used FLASH-based CEST sequence, focusing on SNR and image contrast, while maintaining identical pre-saturation modes, repetition time, echo time and scan time. In the simulation experiments, the maximum CEST signal obtained from the True FISP was significantly greater than that obtained from the FLASH sequence. In the egg white phantom, the SNRs of amide proton transfer-weighted (APTw) and nuclear Overhauser enhancement (NOE) effect images obtained from the True FISP were 68.3% and 57.0% higher than those obtained from the FLASH sequence, respectively. In healthy volunteers, saturated images collected with the True FISP sequence at 3.5 ppm showed an approximate 84% increase in mean temporal SNR compared to those collected with the FLASH sequence. Compared to the FLASH sequence, the CEST images obtained from the True FISP sequence could display more detailed brain tissue structures of both normal individuals and the patient with a brain tumor. Therefore, due to the high SNR inherent in the sequence, True FISP has the potential to be used for fast and high-quality 3D image readout of CEST contrasts in clinical applications.

**Abbreviations used:**

3D, three dimensional;

APT, amide proton transfer;

APTw, amide proton transfer-weighted;

bSSFP, balanced steady state free precession;

B0, main magnetic field;

CEST, chemical exchange saturation transfer;

EPI, echo planar imaging;

FA, flip angle;

FLASH, fast low angle shot;

GM, gray matter;

GRASE, gradient and spin echo;

MRS, magnetic resonance spectroscopy;

MT, magnetization transfer;

NOE, nuclear Overhauser enhancement;

RF, radio frequency;

ROI, region of interest;

SD, standard deviation;

SNR, signal-to-noise ratio;

SPACE, sampling perfection with application of optimized contrasts using different flip angle evolution;

SSFP, steady state free precession;

True FISP, true fast imaging with steady-state precession;

TSE, turbo spin echo;

tSNR, temporal signal-to-noise ratio;

WM, white matter.

## 1 INTRODUCTION

Chemical exchange saturation transfer (CEST) [1-4] is a metabolic MRI technique that enables indirect detection of proteins and metabolites in vivo through exchangeable protons, which are challenging to detect directly by magnetic resonance spectroscopy (MRS) due to their low concentration, exchange-dependent line broadening and/or the effects of water presaturation transferring to solute signals. CEST MRI has shown great promise in various clinical applications, particularly in the diagnosis and evaluation of tumors,[5-7] ischemia,[2,8] and epilepsies.[9,10] Because CEST imaging requires capturing multiple images with different frequency offsets, a very fast image acquisition speed is required to reduce scanning time. In addition, the low signal-to-noise ratio (SNR) of CEST images, caused by scanning time limitations and/or high acceleration acquisition factors, poses a significant challenge in the clinical practice of CEST imaging.[11] This issue is even more pronounced due to the inherent low spectral resolution and imaging SNR at 3T field strength.

The asymmetry analysis of Z-spectrum ($MTR_{asym}$),[2,12] widely used for obtaining CEST effects, requires collecting only a few irradiation frequency offsets (usually within 10) to determine CEST contrasts. This makes it relatively easy to achieve within the limited clinical scanning time. Spectrally selective CEST imaging,[5,13,14] on the other hand, collects the complete Z-spectrum with dozens of irradiation frequency offsets, allowing for multi-pool analysis[13,15] and avoiding aliasing of CEST effects at opposite frequency offset points relative to water. This technique presents multiple CEST contrasts simultaneously, such as amide proton transfer (APT), nuclear Overhauser enhancement (NOE) effect, and other effects from different diluted solutes, each with specific applications.[16-19] However, it has higher requirements for the acquisition speed of image readout, especially in the case of 3D imaging. A variety of 3D rapid imaging sequences, such as gradient and spin echo (GRASE),[20,21] turbo spin echo (TSE)[22,23] and the optimized TSE sequence called sampling perfection with application of optimized contrasts using different flip angle evolution (SPACE),[24] have been employed for asymmetric analysis. It takes about 20-40 seconds for these 3D acquisition sequences to collect a frequency offset image, which is feasible to collect limited irradiation frequency offsets within the clinical scanning time for asymmetric analysis. However, the use of spectrally selective CEST faces challenges due to the extended scanning time required.

Previously, single-shot EPI and fast low-angle shot (FLASH) sequences were used for 3D CEST scanning owing to their fast scanning speeds, enabling the completion of scanning a frequency offset image in 7 seconds. Consequently, acquiring over 50 frequency offset images required less than 6 minutes.[25-27] The single-shot CEST sequence reads out the entire 3D volume after one preparation block, which could significantly reduce the acquisition time. In addition, the single-shot CEST sequence fully separates the preparation block from the readout block, providing the freedom to optimize the CEST preparation block to obtain desired contrasts.[28] This feature of single-shot CEST facilitates the implementation of various spectrum-editing techniques,[29-33] enriching the methods for obtaining different CEST contrasts. However, the rapid imaging of EPI readout is prone to be affected by susceptibility artifacts and lipid-related artifacts, which can potentially affect CEST imaging.[34] Although FLASH enables fast image acquisition, it often results in images with low SNR due to small-angle excitations.[11]

While some works have improved the quality of CEST images and achieved remarkable results through advanced image reconstruction techniques[35-37] and Z-spectrum denoising,[38] the intrinsic SNR from the sequence acquisition process still has a decisive impact on the quality of the final CEST images. A previous study compared the imaging sequences of steady state free precession (SSFP) and FLASH combined with pulsed steady-state CEST.[39] It was observed that SSFP exhibited higher CEST sensitivity than the FLASH sequence. However, it's worth noting that the SSFP sequence used in the previous study was an unbalanced GRE sequence, as pulsed steady-state CEST has a spoiled gradient after each saturation pulse.[40]

True fast imaging with steady-state precession (True FISP) sequence, also known as balanced steady state free precession (bSSFP), is a gradient recalled echo (GRE) variant and has gained increasing importance in various applications.[40] Building on the extended two-pool bSSFP signal model, True FISP has emerged as an effective technique for acquiring high-resolution quantitative MT maps.[41] Furthermore, this methodology has been successfully adapted for chemical exchange detection through exploitation of the off-resonance spectral characteristics inherent to its steady-state signal response (bSSFPX).[42,43] Compared to other

GRE variants, the gradients in the True FISP sequence are fully balanced (gradient moment sums to zero over TR), resulting in more efficient refocusing of the steady-state magnetization and thus increased signals, especially when TR is significantly shorter than T2. Recently, Lukas Vollmann et al.[44] compared the performance of 2D centric-reordered sequences, including True FISP, FLASH, and TSE, in terms of point spread function and SNR through simulations and in-vivo experiments. Their results demonstrated that the SNR of True FISP is comparable to that of TSE, while True FISP outperforms TSE in acquisition speed and sharpness. Moreover, True FISP provides image sharpness comparable to FLASH while offering significantly higher SNR. Therefore, the utilization of 3D True FISP sequence in CEST has great potential to improve the SNR of GRE CEST imaging.

In this study, we developed a single-shot CEST sequence with a 3D True FISP readout to achieve fast 3D CEST imaging with high SNR. Magnetization simulations and MRI experiments were performed to compare the SNR and image quality of CEST imaging between single-shot CEST sequences based on True FISP and FLASH readouts.

## 2 METHODS

### 2.1 Sequence design

The CEST sequence used in this study consists of a pre-saturation module followed by a fast gradient echo readout with a short TR, as shown in Figure 1. At the beginning of the pre-saturation module, a train of 28 Gaussian-shaped RF pulses is applied, each lasting 100 ms with a mean amplitude of 0.6 μT. A crusher gradient is applied during a 5 ms gap between the Gaussian saturation pulses to destroy residual transverse magnetization. At the end of the pre-saturation module, fat saturation is applied, resulting in a total pre-saturation time of $t_{Sat}$ = 3 seconds. Data acquisition is achieved with a 3D GRE sequence using centric spiral reordering.[26] The True FISP readout uses alternating phase RF excitation, with gradients arranged in a way that the net gradient moment in a TR is zero along each axis. To mitigate signal oscillations, a half-flip angle pulse with a half-TR interval is employed as a preparation pulse, that is, $(\theta/2)_x$ – (TR/2) – $\theta_{-x}$ – TR – $\theta_x$ … As a comparison, the CEST sequence based on FLASH uses the same pre-saturation module and readout parameters (TR, TE, readout bandwidth, etc.) as the True FISP sequence.

### 2.2 Magnetization simulation experiments

To investigate whether the True FISP sequence has an increased SNR for CEST imaging compared to the FLASH sequence, the magnetization evolution process of the True FISP readout sequence during the readout of CEST signals was studied and compared with the FLASH sequence according to the scanning parameters: TR, flip angle (FA), and the number of acquired K-space lines.

As shown in Figure 2, at the end of saturation, the CEST effects result in slight differences between the longitudinal magnetization $M_{sat}$ of the CEST label scan (solid red line) and the reference scan (without CEST, dashed yellow line). During the acquisition, the longitudinal magnetization $M_z$ from the initial state $M_{sat}$ gradually reaches a steady state.

The transverse magnetization, which represents the signal read by the True FISP sequence, is dependent on the number of K-space lines collected (n):

$$M_{\perp,n} = \left[M_{\perp,1} + \lambda - \frac{M_{\perp,2}+\lambda-\alpha(M_{\perp,1}+\lambda)}{\beta-\alpha}\right] \cdot \alpha^{n-1} + \frac{M_{\perp,2}+\lambda-\alpha(M_{\perp,1}+\lambda)}{\beta-\alpha} \cdot \beta^{n-1} - \lambda, \quad (1)$$

where,

$E_{1,2} = e^{-TR/T_{1,2}}$,

$M_{\perp,1} = M_{Z,1} \cdot \sin\theta - M_{\perp,0} \cdot \sqrt{E_2} \cdot \cos\theta$,

$M_{\perp,2} = M_{Z,2} \cdot \sin\theta - M_{\perp,1} \cdot E_2 \cdot \cos\theta$,

$M_{Z,0} = M_{sat}$,

$M_{Z,1} = M_{Z,0} \cdot \cos(\theta/2) \cdot \sqrt{E_1} + M_0\left(1 - \sqrt{E_1}\right)$,

$M_{Z,2} = (M_{Z,1} \cdot \cos\theta + M_{\perp,0} \cdot \sqrt{E_2} \cdot \sin\theta) \cdot E_1 + M_0(1 - E_1)$,

$M_{\perp,0} = M_{Z,0} \cdot \sin(\theta/2)$,

$\alpha = \frac{p-\sqrt{p^2+4q}}{2}$,

$\beta = \frac{p+\sqrt{p^2+4q}}{2}$,

$\lambda = \frac{k}{p+q-1}$,

$p = \cos\theta \cdot (E_1 - E_2)$,

$q = E_1 \cdot E_2$,

$k = M_0 \cdot \sin\theta \cdot (1 - E_1)$.

Here, $M_{\perp,1}$ and $M_{\perp,2}$ represent the magnitude of the transverse magnetization after the second and third RF excitations, respectively. $M_{\perp,0}$ is the transverse magnetization after the first RF excitation. The first RF pulse is a $\theta/2$ preparation pulse, and its signal does not fill the k-space. $M_{Z,0}$, $M_{Z,1}$ and $M_{Z,2}$ represent the longitudinal magnetization before the first three RF excitations, respectively. $M_0$ is the longitudinal magnetization after complete $T_1$ recovery. $M_{sat}$ is the longitudinal magnetization value after the pre-saturation module ends, which is also the initial longitudinal magnetization value of the acquisition module. The detailed derivation process of Eq. (1) can be found in the Supporting Information.

The CEST signal can be reasonably modeled as the difference between two signals with different initial longitudinal magnetization value ($M_{sat}$) of the acquisition module.[26] Then, the CEST signal is determined by:

$$\Delta M_{\perp,n} = \Delta M_{sat} \cdot \left[\left(\mu_1 - \frac{\mu_2 - \alpha\cdot\mu_1}{\beta-\alpha}\right) \cdot \alpha^{n-1} + \frac{\mu_2 - \alpha\cdot\mu_1}{\beta-\alpha} \cdot \beta^{n-1}\right], \qquad (2)$$

where,

$\mu_1 = \cos(\theta/2) \cdot \sqrt{E_1} \cdot \sin\theta - \sin(\theta/2) \cdot \sqrt{E_2} \cdot \cos\theta$,

$\mu_2 = \left[\cos(\theta/2) \cdot \sqrt{E_1} \cdot \cos\theta + \sin(\theta/2) \cdot \sqrt{E_2} \cdot \sin\theta\right] \cdot E_1 \cdot \sin\theta - \mu_1 \cdot E_2 \cdot \cos\theta$.

Here, $\Delta M_{sat}$ is the initial longitudinal magnetization difference of acquisition module between the two signals. It also signifies the longitudinal magnetization difference at the end of the pre-saturation module.

The integration of Eq. (2) from n = 1 to n = N generates the total CEST signal achievable from the first N K-space lines collected by True FISP sequence:

$$\Delta S = \Delta M_{sat} \cdot \left[\left(\mu_1 - \frac{\mu_2 - \alpha \cdot \mu_1}{\beta - \alpha}\right) \cdot \frac{(\alpha^{N-1} - 1)}{\ln \alpha} + \frac{\mu_2 - \alpha \cdot \mu_1}{\beta - \alpha} \cdot \frac{(\beta^{N-1} - 1)}{\ln \beta}\right]. \quad (3)$$

From Eq. (3), it can be concluded that for a longitudinal magnetization difference $\Delta M_{sat}$ at the end of the pre-saturation module, the integrated CEST signal ($\Delta S$) is proportional to the subsequent multiplier factor. The scanning parameters, including TR, flip angle, and the number of collected K-space lines (N), were subsequently investigated to maximize the integrated CEST signal.

To investigate whether the True FISP sequence can yield a larger integrated CEST signal ($\Delta S$) than FLASH sequence under the same pre-saturation conditions, similar simulation experiments on FLASH sequence were also conducted for comparison, with specific simulation methods referring to the previous study by Moritz Zaiss et al.[26] Detailed formula derivation can be found in the Supporting Information.

The simulation experiments simulated the magnetization of brain white matter ($T_1$ = 1087 ms, $T_2$ = 56 ms) and gray matter ($T_1$ = 1703 ms, $T_2$ = 71 ms) at 3T.[45,46]

### 2.3 MRI experiments

MR imaging was performed on a 3T whole-body MRI system (MAGNETOM Prisma Fit; Siemens Healthcare, Erlangen, Germany) on an egg white phantom and ten healthy volunteers. Additionally, two patients with brain tumors (glioblastoma, WHO grade IV) were scanned using a separate 3T whole-body MRI system (MAGNETOM Prisma; Siemens Healthcare). Egg white, which is the cytoplasm of egg, is a gel-like translucent liquid mixture containing about 11% protein. It exhibits strong APT and NOE effects in MRI experiments, and is an ideal simple model.[47,48] The egg white phantom was created by placing a cylindrical plastic bottle filled with egg white (pH = 7.8) into a larger diameter cylindrical plastic container filled with water. The vendor's Head/Neck 64-channel coil was used for signal reception. This study was approved by the local institutional review board, and written informed consent was obtained from each participant.

For the CEST imaging of both the egg white phantom and human subjects, the scan parameters were set as follows: TR = 4 ms, echo time (TE) = 2 ms, FA = 5° (FLASH sequence) or 26° (True FISP sequence), bandwidth = 550 Hz/pixel, field of view = 220 × 180 × 90 $mm^3$, matrix size = 128 × 104 × 18, voxel size = 1.7 × 1.7 × 5 $mm^3$. A thirteen-lobe sinc waveform of the excitation RF pulse was used for slab-selective excitation to reduce the aliasing artifact in the slice-selection direction. A 6/8 partial Fourier acquisition was employed in both phase and slice encoding directions, and elliptical sampling was used. These resulted in a readout time of $t_{RO}$ = 3.6 seconds. The acquisition time of each frequency offset was TA = $t_{Sat}$+$t_{RO}$ = 6.6 seconds. Z-spectrum data were obtained after saturation at 55 irradiation frequency offsets: two at –300 ppm for unsaturated reference images, ±50 ppm, ±35 ppm, ±20 to ±11 ppm in steps of 3 ppm, and -10 to 10 ppm in steps of 0.5 ppm. The total scanning time of the Z-spectrum, with 55 irradiation frequency offsets, was approximately 6.1 minutes.

To evaluate the SNRs of CEST maps acquired with two different readout sequences, the egg white phantom was scanned twice consecutively using the same sequence. Additionally, to compare the temporal signal-to-noise ratios (tSNRs) of the two distinct readout sequences, a healthy volunteer underwent 20 continuous scans, maintaining a saturation frequency offset of 3.5 ppm.

Furthermore, high-resolution $T_1$-weighted images of the whole brain were acquired from each healthy volunteer by using magnetization prepared rapid gradient echo imaging (MP-RAGE) sequence.

For the comparative evaluation of two CEST sequences in brain tumor assessment, one patient with brain tumor underwent paired scanning using True FISP and FLASH sequences. Conventional MR imaging protocols - including axial $T_1$-weighted, $T_2$-weighted, and contrast-enhanced $T_1$-weighted sequences - were systematically acquired to establish comprehensive anatomical references.

To further evaluate the reliability of the True FISP sequence for CEST applications within the context of the existing clinical APTw literature, supplementary APTw imaging was performed on a second brain tumor patient. The APTw saturation parameters strictly followed established consensus guidelines[6], employing a saturation power of 2 μT saturation power and a saturation duration of 2 seconds. Frequency-offset acquisitions were conducted at ±3 ppm, ±3.5 ppm, and ±4 ppm for the saturated images, with an additional acquisition at -300 ppm to obtain unsaturated reference images. Each acquisition was repeated three times to improve SNR. The multi-pool CEST protocol proposed was simultaneously implemented using the True FISP readout.

### 2.4 Data analysis

Data processing codes were developed using the MatLab R2023a platform (The MathWorks, Inc., Natick, MA, USA).

The Z value of the Z-spectrum at RF frequency offset Δω was given by the saturated image $S_{sat}(\Delta\omega)$ and the unsaturated image $S_0$, as described by the following equation:

$$Z(\Delta\omega) = \frac{S_{sat}(\Delta\omega)}{S_0}. \quad (4)$$

For B0 correction, the Z-spectrum was adjusted by interpolating the raw Z-spectrum to a frequency interval of 1 Hz and shifting it along the offset axis to align with its lowest intensity at 0 ppm.

The four-pool Lorentz fitting[15] method was used to process the Z-spectrum voxel by voxel from both the egg white phantom and the human brain, including amide (+3.5 ppm), NOE (−3.5 ppm), MT (−1.0 ppm), and water (0 ppm). Eq. (5) provides the model function for the Lorentz fitting method:

$$Z(\Delta\omega) = 1 - \sum_{i=1}^{4} L_i(\Delta\omega), \quad (5)$$

where,

$$L_i(\Delta\omega) = A_i \frac{{W_i}^2/4}{{W_i}^2/4 + (\Delta\omega - \Delta_i)^2}.$$

Here, $A_i$, $\Delta_i$, and $W_i$ are the amplitude, center frequency, and full width at half maximum of the ith Lorentzian pool ($i$ = 1, 2, 3, 4). The fitting parameters are shown in Supporting Information Table S1.

The spillover and MT-corrected inverse difference (Rex) method[49,50] was used to evaluate the CEST effects:

$$MTR_{Rex}(\Delta\omega) = \frac{1}{Z_{lab}(\Delta\omega)} - \frac{1}{Z_{ref}(\Delta\omega)}. \quad (6)$$

Here, $MTR_{Rex}$ provides a spillover and MT-corrected CEST contrast for the exchange-dependent relaxation contribution; $Z_{ref}$ denotes the reference signals obtained by summing all Lorentzian pools except the corresponding pool estimated; $Z_{lab}$ represents the signals of the four fitted Lorentzian pools.

The SNRs of the CEST maps of the egg white phantom were determined by the following equation:[51]

$$\mathrm{SNR} = \sqrt{2} \cdot \frac{\overline{MTR_{Rex}(\Delta\omega)}}{SD}. \quad (7)$$

Here, $\overline{MTR_{Rex}(\Delta\omega)}$ represents the average intensity value of voxels in the region of interest (ROI) across 10 slices containing CEST signal from the first scan. Due to the limited length of the egg white phantom, only 10 slices in the axial direction contained CEST signals from the egg white. Therefore, the SNR was calculated from these 10 slices. SD is the standard deviation within the ROI of the difference images between the first and second scans.

The contrast-to-noise ratios (CNRs) of CEST images obtained from the first scan by True FISP and FLASH sequences were evaluated. The egg white region was designated as the target area, while the surrounding water served as the background. The CNR was calculated using the following formula:

$$\mathrm{CNR}(\Delta\omega) = \frac{\overline{MTR_{Rex}(\Delta\omega)}_{target} - \overline{MTR_{Rex}(\Delta\omega)}_{\mathrm{background}}}{\mathrm{SD}(\Delta\omega)_{\mathrm{background}}}. \quad (8)$$

Here, $\overline{MTR_{Rex}(\Delta\omega)}_{target}$ and $\overline{MTR_{Rex}(\Delta\omega)}_{\mathrm{background}}$ represent the average intensity value of voxels in the egg white region and the surrounding water region, respectively. $\mathrm{SD}(\Delta\omega)_{background}$ is the standard deviation within the surrounding water region.

Twenty consecutive brain CEST images from a healthy volunteer, obtained at a saturation RF frequency offset of 3.5 ppm, were used to calculate tSNR with the following equation:[25]

$$tSNR = \frac{\overline{S_{sat}(3.5\mathrm{ppm})}}{SD}. \tag{9}$$

Here, $\overline{S_{sat}(3.5\mathrm{ppm})}$ is the average signal intensity, and SD is the standard deviation of 20 scans at the same location. High-resolution $T_1$-weighted images were then aligned to CEST images by registering them with the unsaturated images acquired during CEST imaging.

For the patient with a brain tumor, all conventional MR images were also aligned to CEST images by registering with the unsaturated images of CEST acquisition. The software SPM12 (v7771) was used for both segmentation and registration processes. The ROI of solid tumor was defined as the enhanced region in the contrast-enhanced $T_1$-weighted image (Figure 8A). Additionally, the ROI of normal appearing white matter on the contralateral hemisphere was drawn on the $T_1$-weighted image (Figure 8B). The averages and standard deviations of the CEST signals in the ROIs were measured.

## 3 RESULTS

### 3.1 Contrast and imaging optimization

Figure 3 shows the results of magnetization simulation experiments of GRE readout in white matter, with both $M_0$ and $\Delta M_{sat}$ normalized. For both FLASH and True FISP sequences, a shorter TR is beneficial to obtain a larger integrated CEST signal for N = ∞ (Figure 3A). Taking into account the duration of the selective excitation pulse and the sampling bandwidth, a TR of 4 ms was selected for the subsequent calculations. For N = ∞, the maximum integrated CEST signal can be obtained near a flip angle of 5° for the FLASH sequence and 26° for the True FISP sequence, respectively (Figure 3B). Moreover, the highest integrated CEST signal value achieved by the True FISP sequence is much larger than that obtained by FLASH. With limited N values, the flip angle corresponding to the peak of the integrated CEST signal increases. However, the aforementioned conclusion remains valid for sufficiently large N values. Considering that the 3D GRE CEST acquisition sequence usually requires the acquisition of over 500 K-space lines to achieve adequate resolution and SNR,[26,27] we chose 5° and 26° as the optimal flip angles for FLASH and True FISP, respectively, for subsequent simulations.

Figure 3C shows the attenuation process of the CEST signal ($\Delta M_{\perp}$) with the number of K-space lines collected by the two GRE sequences under the condition of TR = 4 ms and flip angles of 5° for FLASH and 26° for True FISP. Throughout the entire acquisition process, True FISP maintains a higher CEST signal than FLASH. Figure 3D illustrates the evolution of transverse magnetization of FLASH and True FISP sequences under different initial longitudinal magnetization values $M_{sat}$ with the optimized imaging parameters mentioned above. It demonstrates that the signal of True FISP is much larger than FLASH in both unsaturated and saturated situations. The simulation experiment of gray matter in the brain yields results similar to those of white matter, as shown in Supporting Information Figure S1.

### 3.2 Egg white phantom

Figure 4 shows unsaturated images, APTw, NOE, MT maps and examples of fitted Z-spectrum of egg white phantom acquired using True FISP and FLASH sequences, respectively. Notably, CEST images obtained from both sequences exhibited similar contrasts. The egg white demonstrated strong APTw and NOE effects, with negligible MT effect. The water surrounding the egg white container had a value of 0 in the CEST images acquired by True FISP and FLASH sequences. The fitting residuals of the Z-spectrum for both sequences were less than 1%. No significant blurring effects induced by point spread function were observed in the unsaturated images of True FISP and FLASH sequences. However, the CEST images obtained from the FLASH sequence exhibit noticeable blurring in the phase-encoding direction (indicated by the red arrow).

Table 1 shows the SNRs of APTw and NOE maps of the egg white phantom acquired using True FISP and FLASH sequences. Compared to the FLASH sequence, the SNRs of the APTw and NOE maps with the True FISP sequence increased by 68.3% and 57.0%, respectively.

Table 2 shows the CNRs of APTw and NOE maps of the egg white phantom acquired using True FISP and FLASH sequences. Compared to the FLASH sequence, the CNRs of the APTw and NOE maps with the True FISP sequence increased by 134.4% and 182.8%, respectively.

### 3.3 Healthy volunteers

Figure 5A shows unsaturated images and tSNRs of saturated images at 3.5 ppm of a healthy volunteer's brain, acquired using True FISP and FLASH sequences on the same slice. Figure 5B shows the mean tSNRs in the axial slices obtained from True FISP and FLASH sequences. The mean tSNR was calculated from the brain region after removing extracerebral tissue. The True FISP sequence exhibited an approximately 84% increase in mean tSNR compared to the FLASH sequence.

Figure 6 shows the APTw, NOE, MT maps and examples of fitted Z-spectrum in the occipital lobe of a healthy subject's brain acquired using the True FISP and the FLASH sequences. The fitting residuals of the Z-spectrum for both sequences were less than 1.5%. It was observed that the True FISP readout could detect finer anatomical structures in the brain compared to FLASH readout. In particular, the fitted MT images appeared blurrier when using the FLASH readout. All slices of the CEST images, along with their corresponding tSNR and $\Delta B0$ images, are presented in Supporting Information Figure S2. A review of all slices of unsaturated images and fitted CEST images showed no sign of strip-shaped banding artifacts.

Figure 7 shows the statistical data of the average $MTR_{Rex}$ values in the WM and GM regions across ten healthy volunteers. There were no significant differences in the mean $MTR_{Rex}$ values of brain tissues between the True FISP and FLASH sequences (all $P$ values > 0.05).

The average SARs during the measurement process across the ten subjects were 62.3% ± 7.8% for the True FISP sequence and 6.0% ± 1.0% for the FLASH sequence.

### 3.4 Patients with brain tumors

Figure 8 shows the conventional MRI images along with fitted CEST images from the True FISP sequence and the FLASH sequence of a patient with a brain tumor. The CEST images obtained from both sequences showed similar contrast between tumor tissue and normal brain tissue. However, compared to the FLASH sequence, the CEST images obtained from the True FISP sequence revealed a clearer boundary of tumor tissue, while the CEST images from the FLASH sequence presented a more blurred appearance.

A notable decrease in APTw and NOE signals was observed in the tumor necrosis area. Conversely, in the solid tumor area, there was an increase in APTw signal and a decrease in NOE signal. Table 3 presents the CEST values from two sequences in the solid tumor region and the contralateral normal white matter region in the patient. The CEST values for the True FISP sequence and FLASH sequence are similar in both the tumor area and the contralateral normal white matter areas.

Figure 9 presents APTw (2 uT) and multi-pool CEST images of the second brain tumor patient acquired using the True FISP readout, along with corresponding conventional MR images. The APTw (2 uT) images clearly exhibit hyperintensity in tumor region, with near-zero background signal in surrounding normal brain tissue.

## 4 DISCUSSION

In this study, we developed a single-shot, spectrally selective CEST sequence based on True FISP readout to achieve fast and high-SNR 3D CEST imaging. Through mathematical derivation, we elucidated the physical mechanism of signal acquisition in the GRE-CEST sequences and optimized the acquisition parameters through simulation experiments. The egg white phantom and in vivo experiments demonstrated that the True FISP sequence significantly improved the SNR of CEST imaging compared to the commonly used FLASH based 3D single-shot CEST sequence, without increasing time costs and while maintaining the same spatial resolution.

In the egg white phantom experiments, the SNRs of the CEST maps obtained with the True FISP sequence exhibited a significant improvement (approximately 60%) compared to the FLASH sequence. A higher tSNR was observed in human experiments for saturated images collected with True FISP sequence at 3.5 ppm compared to FLASH sequence. Additionally, the human brain CEST maps acquired with True FISP sequence revealed finer anatomical structures and could more clearly display the boundaries between tumor area and normal brain tissues. These findings can be explained by the results of simulation experiments (Figure 3). With their respective optimal flip angles, True FISP sequence can obtain a larger integrated CEST signal ($\Delta S$) than FLASH sequence for the same CEST saturation ($\Delta M_{sat}$), leading to increased SNRs of CEST images (Figure 3B); the True FISP sequence can also obtain larger K-space signals than those collected with FLASH in both saturated and unsaturated states, leading to increased tSNRs of saturated images (Figure 3D).

A higher SNR of True FISP readout makes CEST images more accurate and displays clear tissue contrast. While unsaturated images of the egg white phantom acquired with both True FISP and FLASH sequences show comparable sharpness, CEST images generated using the FLASH sequence exhibit significant blurring in the phase-encoding direction (Figure 4), contrasting with the clarity of True FISP-derived CEST images. This difference cannot be attributed solely to the point spread function (PSF) influencing the raw signal decay (governing $M_{\perp,n}$ in Figure 3D) during acquisition, as evidenced by the sharp unsaturated images from both sequences. Instead, the blurring in FLASH CEST images arises from the characteristics of the CEST signal dynamics during k-space filling. Here, the "CEST signal" specifically denotes the CEST-induced signal change ($\Delta M_{\perp,n}$, Figure 3C), which determines contrast in CEST images (Equation 2). Crucially, both sequences exhibit similar temporal CEST signal decay patterns during k-space line acquisition (Figure 3C). However, due to center-out k-space sampling, the CEST signal attenuates towards the periphery of k-space, potentially reaching the noise floor (where steady-state $\Delta M_{\perp,n} \approx 0$). Since high spatial frequency information resides in these outer k-space regions, signal loss there disproportionately degrades resolution. Under identical acquisition parameters (k-space line count and trajectory), True FISP maintains significantly higher CEST signal intensity than FLASH throughout k-space, both centrally and peripherally. This superior signal preservation in True FISP delays the signal decay towards the noise floor, thereby better retaining critical high-frequency spatial information and mitigating phase-encoding blurring in the resulting CEST images.

An acquisition sequence with a higher SNR not only provides more accurate CEST images,

but also plays a significant role in improving the number of slices and spatial resolution, particularly for the single-shot sequence discussed here. In single-shot CEST imaging, maintaining the contrast and resolution of CEST images is essential, as the attenuation of CEST contrast during the acquisition limits the number of collectible K-space lines after a single pre-saturation module (Figure 3C). In comparison to FLASH, the inherently higher SNR of CEST imaging in the True FISP sequence has the potential to achieve a larger spatial coverage and a higher spatial resolution with a greater acceleration factor, while still providing reliable CEST images. This is of great significance for clinical diagnosis, especially in scenarios involving multiple lesions at different brain locations or when identifying minor lesions.[52]

The values of CEST images collected using the True FISP and the FLASH sequences remained generally consistent, as confirmed by the experimental results from both healthy individuals and a patient with a brain tumor. This suggests that the True FISP sequence provides a higher SNR for CEST images than the FLASH sequence, without altering the values of CEST effects. In this study, both the True FISP and FLASH sequences used identical pre-saturation modules, ensuring consistent initial magnetization before signal acquisition. Additionally, both sequences employed 3D acquisition with centric spiral reordering, where the K-space center was acquired first. Consequently, the CEST contrast at the K-space center in both sequences was identical, reflecting the identical magnetization state at the end of the pre-saturation module. Since the K-space center determines the low-frequency information (i.e., image contrast), the CEST contrast remains the same in both sequences. Additionally, APTw (2 uT) imaging in the second tumor patient demonstrated tumor hyperintensity and near-zero background signal in surrounding normal brain tissue, consistent with characteristic patterns reported in clinical APTw literature.[6]

In the single-shot CEST method, the separation of saturation and acquisition processes allows for independent optimization of these two modules. According to Eq. (3), the CEST signal obtained during the readout stage is proportional to the longitudinal magnetization difference ($\Delta M_{sat}$) at the end of the saturation stage. Therefore, the situation during the saturation stage does not affect the parameter optimization for the readout stage. Following the scheme to optimize saturation module in CEST imaging proposed by Deshmane et al.,[27] $B_{1,mean}$ was set to 0.6 uT in this study to achieve both NOE effect and APTw contrasts simultaneously.

In the CEST images of the patient with a brain tumor acquired from both sequences, the solid tumor demonstrated notable contrast differences compared to the contralateral normal white matter region. Specifically, there was a marked increase in the APTw signal, a subtle reduction in the NOE signal, and a substantial decrease in the MT signal within the solid tumor area. However, for a conclusive assessment of the observed contrast changes to clinical questions, a sufficiently large cohort needs to be investigated.

It has been noticed that distinct quantitative metrics in CEST imaging can exert a substantial influence on CEST contrast, for instance, in the contrast between tumors and healthy brain tissue. APT effect primarily arises from the saturation-transfer effect of mobile proteins/peptides.[2] A recent study has demonstrated that the NOE at -3.5 ppm in the brain predominantly originates from membrane lipids.[53] APTw (2 uT), based on symmetry analysis, showed an increase in APTw effect in tumors relative to normal tissues, which was attributed to the proliferation of tumor cells.[6,12] The Lorentzian difference ($MTR_{LD}$) revealed an increase in the APTw signal and a slight decrease in the NOE signal in the solid tumor area, while both signals showed hypointensity in the necrosis area compared to normal tissue.[27] $MTR_{Rex}$, the spillover and MT corrected inverse difference used in this study, presents a contrast between the tumor area and normal tissue similar to that of $MTR_{LD}$.[5]

$T_1$ correction was not carried out in this study because obtaining a 3D or multi-slice $T_1$ map using traditional methods (IR-TSE) requires a substantial amount of scan time. The absence of $T_1$ correction could significantly affect the CEST contrast of tumors, as high $T_1$ values in the tumor area might lead to an overestimation of APTw and NOE signals. Apparent exchange-dependent relaxation (AREX) further corrects for $T_1$ relaxation. Previous studies have indicated that AREX shows similar APTw signal intensity between the tumor area and normal tissue, while enhancing the reduction of the NOE signal in the tumor area.[5] Jing Cui et al.'s research[54] suggests that CEST metrics without $T_1$ correction reflects the combined effects of $T_1$ weighting and APT/NOE. Both of these factors rely on solute concentration and strengthen the correlation of the APT effect with solute concentration in APTw imaging.

Consequently, CEST metrics after $T_1$ correction may not necessarily improve diagnostic effectiveness. In conclusion, the clinical efficacy of CEST metrics with $T_1$ correction still requires further verification, and fast, accurate $T_1$ imaging suitable for CEST correction is a key prerequisite for promoting the application of AREX in routine clinical scans.

Additionally, B1 correction was not applied, which could potentially impact the accuracy of the obtained CEST effect. Although B1 correction is less commonly applied in head CEST imaging at 3T field strength (due to relatively less severe B1 inhomogeneity compared to higher field strengths like 7T),[55] we observed a signal-reduction trend in the frontal lobe region of the MT images (Figure 6). This is likely due to the lower B1 amplitude in that region. To improve the accuracy and reproducibility of CEST imaging, it is recommended to use B1 field correction in future clinical scans.

This study employs a four-pool Lorentz fitting approach, which may cause the APT effect to overlap with fast exchange effects at nearby frequency points. Recent research suggests that even at low saturation powers, the fast exchange associated with the amine CEST effect can result in overestimation of the APT effect, amide concentration, and amide-water exchange rate, potentially causing inconsistencies in previous research results.[56] For more accurate quantification of APT, amine, and guanidine, specific Z-spectrum analysis methods such as polynomial and Lorentzian line-shape fitting (PLOF)[57] and double saturation power techniques[58] should be considered.

The True FISP sequence is associated with higher SAR values, mainly due to the larger flip angle needed for optimal signal acquisition. In our study, the True FISP sequence had a 26-degree flip angle, resulting in a mean SAR value of 62%, significantly higher than the 6% SAR of the FLASH sequence with a 5-degree flip angle. This elevated SAR is troublesome, especially for detecting exchangeable protons with fast exchange rates, as they require higher saturation powers, further worsening the SAR. For future studies, we suggest using asymmetric RF pulses, which can reduce RF bandwidth and thereby lower SAR values.

True FISP is sensitive to the B0 inhomogeneity and regional magnetic susceptibility differences, which can lead to banding artifacts.[40] This sensitivity may pose a greater challenge in body imaging compared to head imaging, as the issues are likely to be more pronounced. To mitigate or eliminate the influence of banding artifacts in CEST imaging with single-shot True FISP, the following strategies can be considered: (1) perform carefully active shimming before applying the True FISP sequence to ensure the B0 field homogeneity in the target area; (2) set TR to the shortest possible value, as reducing TR enhances tolerance to off-resonance effects without introducing banding artifacts; (3) apply fat-suppression techniques, such as incorporating a fat-saturation pulse before signal readout, to reduce the off-resonance effects caused by chemical shift, which has been included in our proposed sequence; and (4) use specialized True FISP sequences like constructive interference in the steady state (CISS).[59,60] Furthermore, True FISP sequences remain intrinsically sensitive to B0 field inhomogeneity. Head motion during scanning may perturb local B0 distributions, potentially introducing signal instability and reducing tSNR. These effects could generate motion-dependent artifacts or noise patterns specific to True FISP, even in the absence of visible banding artifacts. Future studies should systematically evaluate and mitigate the motion sensitivity of True FISP-based CEST sequences to support robust clinical translation.

Various post-processing algorithms can also greatly improve the quality and speed of CEST imaging, such as advanced K-space-based reconstruction technology,[37,61] motion correction,[62] and Z-spectrum denoising.[38,63] Our next objective is to improve resolution and achieve whole-brain CEST imaging with high SNR based on the True FISP sequence by applying advanced post-processing algorithms.

## 5 CONCLUSIONS

3D single-shot CEST imaging with True FISP readout can be performed at 3T in 6.1 minutes. The True FISP sequence can achieve CEST imaging with a higher SNR and display more detailed brain tissue structures of both normal individuals and tumor patients compared to FLASH under the same spatiotemporal conditions. Therefore, True FISP has the potential to be used for fast and high-quality 3D image readout of CEST contrasts in clinical applications.

## ACKNOWLEDGMENTS

This study was supported by the STI 2030—Major Projects (No. 2021ZD0200500) and Microscale Magnetic Resonance Platform of East China Normal University.

## DATA AVAILABILITY STATEMENT

The data that support the findings of this study are available from the corresponding author upon reasonable request.

## REFERENCES

1. Ward KM, Aletras AH, Balaban RS. A new class of contrast agents for MRI based on proton chemical exchange dependent saturation transfer (CEST). J Magn Reson. 2000;143:79-87.
2. Zhou JY, Payen JF, Wilson DA, Traystman RJ, van Zijl PCM. Using the amide proton signals of intracellular proteins and peptides to detect pH effects in MRI. Nat Med. 2003;9:1085-1090.
3. van Zijl PCM, Yadav NN. Chemical exchange saturation transfer (CEST): What is in a name and what isn't? Magn Reson Med. 2011;65:927-948.
4. Ju LC, Wang KX, Schär M, et al. Simultaneous creatine and phosphocreatine mapping of skeletal muscle by CEST MRI at 3T. Magn Reson Med. 2023;91:942-954.
5. Zaiss M, Windschuh J, Paech D, et al. Relaxation-compensated CEST-MRI of the human brain at 7T: Unbiased insight into NOE and amide signal changes in human glioblastoma. Neuroimage. 2015;112:180-188.
6. Zhou JY, Zaiss M, Knutsson L, et al. Review and consensus recommendations on clinical APT-weighted imaging approaches at 3T: Application to brain tumors. Magn Reson Med. 2022;88:546-574.
7. Koike H, Morikawa M, Ishimaru H, Ideguchi R, Uetani M, Miyoshi M. Amide proton transfer-chemical exchange saturation transfer imaging of intracranial brain tumors and tumor-like lesions: Our experience and a review. Diagnostics. 2023;13:914.
8. Longo DL, Cutrin JC, Michelotti F, Irrera P, Aime S. Noninvasive evaluation of renal pH homeostasis after ischemia reperfusion injury by CEST-MRI. NMR Biomed. 2017;30.
9. Davis KA, Nanga RPR, Das S, et al. Glutamate imaging (GluCEST) lateralizes epileptic foci in nonlesional temporal lobe epilepsy. Sci Transl Med. 2015;7:309ra161.
10. Wen QQ, Wang K, Hsu YC, et al. Chemical exchange saturation transfer imaging for epilepsy secondary to tuberous sclerosis complex at 3T: Optimization and analysis. NMR Biomed. 2021;34:e4563.
11. Zhang Y, Zu T, Liu RB, Zhou JY. Acquisition sequences and reconstruction methods for fast chemical exchange saturation transfer imaging. NMR Biomed. 2022;36:e4699.
12. Zhou JY, Heo HY, Knutsson L, van Zijl PCM, Jiang SS. APT-weighted MRI: Techniques, current neuro applications, and challenging issues. J Magn Reson Imaging. 2019;50:347-364.
13. Heo HY, Zhang Y, Jiang SS, Lee DH, Zhou JY. Quantitative assessment of amide proton transfer (APT) and nuclear Overhauser enhancement (NOE) imaging with extrapolated semisolid magnetization transfer reference (EMR) signals: II. Comparison of three EMR models and application to human brain glioma at 3 Tesla. Magn Reson Med. 2016;75:1630-1639.

14. Desmond KL, Mehrabian H, Chavez S, et al. Chemical exchange saturation transfer for predicting response to stereotactic radiosurgery in human brain metastasis. Magn Reson Med. 2017;78:1110-1120.

15. Zaiss M, Schmitt B, Bachert P. Quantitative separation of CEST effect from magnetization transfer and spillover effects by Lorentzian-line-fit analysis of z-spectra. J Magn Reson. 2011;211:149-155.

16. Haris M, Cai KJ, Singh A, Hariharan H, Reddy R. In vivo mapping of brain myo-inositol. Neuroimage. 2011;54:2079-2085.

17. Haris M, Singh A, Cai K, et al. Micest: A potential tool for non-invasive detection of molecular changes in Alzheimer's disease. J Neurosci Methods. 2013;212:87-93.

18. Haris M, Singh A, Cai KJ, et al. A technique for in vivo mapping of myocardial creatine kinase metabolism. Nat Med. 2014;20:209-214.

19. Cai KJ, Haris M, Singh A, et al. Magnetic resonance imaging of glutamate. Nat Med. 2012;18:302-306.

20. Zhou JY, Zhu H, Lim M, et al. Three-dimensional amide proton transfer MR imaging of gliomas: Initial experience and comparison with gadolinium enhancement. J Magn Reson Imaging. 2013;38:1119-1128.

21. Zhu H, Jones CK, van Zijl PCM, Barker PB, Zhou JY. Fast 3D chemical exchange saturation transfer (CEST) imaging of the human brain. Magn Reson Med. 2010;64:638-644.

22. Zhao XN, Wen ZB, Zhang G, et al. Three-dimensional turbo-spin-echo amide proton transfer MR imaging at 3-Tesla and its application to high-grade human brain tumors. Molecular Imaging and Biology. 2013;15:114-122.

23. Togao O, Keupp J, Hiwatashi A, et al. Amide proton transfer imaging of brain tumors using a self-corrected 3D fast spin-echo Dixon method: Comparison with separate b-0 correction. Magn Reson Med. 2017;77:2272-2279.

24. Zhang Y, Yong XW, Liu RB, et al. Whole-brain chemical exchange saturation transfer imaging with optimized turbo spin echo readout. Magn Reson Med. 2020;84:1161-1172.

25. Mueller S, Stirnberg R, Akbey S, et al. Whole brain snapshot CEST at 3T using 3D-EPI: Aiming for speed, volume, and homogeneity. Magn Reson Med. 2020;84:2469-2483.

26. Zaiss M, Ehses P, Scheffler K. Snapshot-CEST: Optimizing spiral-centric-reordered gradient echo acquisition for fast and robust 3D CEST MRI at 9.4T. NMR Biomed. 2018;31:e3879.

27. Deshmane A, Zaiss M, Lindig T, et al. 3D gradient echo snapshot CEST MRI with low power saturation for human studies at 3T. Magn Reson Med. 2019;81:2412-2423.

28. Sedykh M, Liebig P, Herz K, et al. Snapshot CEST plus plus : Advancing rapid whole-brain APTw-CEST MRI at 3 T. NMR Biomed. 2023:e4955.

29. van Zijl P, Sehgal A. Proton chemical exchange saturation transfer (CEST) MRS and MRI. eMagRes. 2016;5.

30. Zu ZL, Janve VA, Li K, Does MD, Gore JC, Gochberg DF. Multi-angle ratiometric approach to measure chemical exchange in amide proton transfer imaging. Magn Reson Med. 2012;68:711-719.

31. Song XL, Gilad AA, Joel S, et al. CEST phase mapping using a length and offset varied saturation (LOVARS) scheme. Magn Reson Med. 2012;68:1074-1086.

32. Scheidegger R, Vinogradov E, Alsop DC. Amide proton transfer imaging with improved robustness to magnetic field inhomogeneity and magnetization transfer asymmetry using saturation with frequency alternating RF irradiation. Magn Reson Med. 2011;66:1275-1285.

33. Friedman JI, Xia D, Regatte RR, Jerschow A. Transfer rate edited experiment for the selective detection of chemical exchange via saturation transfer (TRE-CEST). J Magn Reson. 2015;256:43-51.
34. Sun PZ, Zhou JY, Sun WY, Huang J, van Zijl PCM. Suppression of lipid artifacts in amide proton transfer imaging. Magn Reson Med. 2005;54:222-225.
35. Heo HY, Zhang Y, Lee DH, Jiang SS, Zhao XN, Zhou JY. Accelerating chemical exchange saturation transfer (CEST) MRI by combining compressed sensing and sensitivity encoding techniques. Magn Reson Med. 2017;77:779-786.
36. Guo CL, Wu J, Rosenberg JT, Roussel T, Cai SH, Cai CB. Fast chemical exchange saturation transfer imaging based on propeller acquisition and deep neural network reconstruction. Magn Reson Med. 2020;84:3192-3205.
37. Xu JP, Zu T, Hsu YC, Wang XL, Chan KWY, Zhang Y. Accelerating CEST imaging using a model-based deep neural network with synthetic training data. Magn Reson Med. 2023;91:583-599.
38. Breitling J, Deshmane A, Goerke S, et al. Adaptive denoising for chemical exchange saturation transfer MR imaging. NMR Biomed. 2019;32:e4133.
39. Kim B, So S, Park H. Optimization of steady-state pulsed CEST imaging for amide proton transfer at 3T MRI. Magn Reson Med. 2019;81:3616-3627.
40. Markl M, Leupold J. Gradient echo imaging. J Magn Reson Imaging. 2012;35:1274-1289.
41. Gloor M, Scheffler K, Bieri O. Quantitative magnetization transfer imaging using balanced SSFP. Magn Reson Med. 2008;60:691-700.
42. Zhang S, Liu Z, Grant A, Keupp J, Lenkinski RE, Vinogradov E. Balanced steady-state free precession (bSSFP) from an effective field perspective: Application to the detection of chemical exchange (bSSFPX). J Magn Reson. 2017;275:55-67.
43. Heule R, Deshmane A, Zaiss M, Herz K, Ehses P, Scheffler K. Structure or exchange? On the feasibility of chemical exchange detection with balanced steady-state free precession in tissue - an in vitro study. NMR Biomed. 2020;33.
44. Vollmann. L, Weinmüller. S, Freudensprung. M, et al. Analysis of TSE, FLASH and bSSFP identifies bSSFP as promising snapshot CEST readout. ESMRMB. 2024;0483.
45. Cheng HLM, Wright GA. Rapid high-resolution T1 mapping by variable flip angles: Accurate and precise measurements in the presence of radiofrequency field inhomogeneity. Magn Reson Med. 2006;55:566-574.
46. Gelman N, Gorell JM, Barker PB, et al. MR imaging of human brain at 3.0 T: Preliminary report on transverse relaxation rates and relation to estimated iron content. Radiology. 1999;210:759-767.
47. Zhou JY, Hong XH, Zhao XN, Gao JH, Yuan J. APT-weighted and NOE-weighted image contrasts in glioma with different RF saturation powers based on magnetization transfer ratio asymmetry analyses. Magn Reson Med. 2013;70:320-327.
48. Lu JH, Zhou JY, Cai CB, Cai SH, Chen Z. Observation of true and pseudo NOE signals using CEST-MRI and CEST-MRS sequences with and without lipid suppression. Magn Reson Med. 2015;73:1615-1622.
49. Zaiss M, Bachert P. Exchange-dependent relaxation in the rotating frame for slow and intermediate exchange - modeling off-resonant spin-lock and chemical exchange saturation transfer. NMR Biomed. 2013;26:507-518.
50. Zaiss M, Windschuh J, Paech D, et al. Relaxation-compensated CEST-MRI of the human brain at 7

T: Unbiased insight into NOE and amide signal changes in human glioblastoma. Neuroimage. 2015;112:180-188.

51. Association NEM. Determination of signal-to-noise ratio and image uniformity for single-channel, non-volume coils in diagnostic magnetic resonance imaging. Arlington:NEMA. 2021:1-21.
52. Dula AN, Asche EM, Landman BA, et al. Development of chemical exchange saturation transfer at 7T. Magn Reson Med. 2011;66:831-838.
53. Zhao Y, Sun CS, Zu ZL. Assignment of molecular origins of NOE signal at-3.5 ppm in the brain. Magn Reson Med. 2023;90:673-685.
54. Cui J, Zhao Y, Sun C, Xu JZ, Zu ZL. Evaluation of contributors to amide proton transfer-weighted imaging and nuclear Overhauser enhancement-weighted imaging contrast in tumors at a high magnetic field. Magn Reson Med. 2023;90:596-614.
55. Hunger L, Rajput JR, Klein K, et al. Deepcest 7 T: Fast and homogeneous mapping of 7 T CEST MRI parameters and their uncertainty quantification. Magn Reson Med. 2023;89:1543-1556.
56. Sun C, Zhao Y, Zu ZL. Validation of the presence of fast exchanging amine CEST effect at low saturation powers and its influence on the quantification of APT. Magn Reson Med. 2023;90:1502-1517.
57. Ju LC, Wang KX, Schär M, et al. Simultaneous creatine and phosphocreatine mapping of skeletal muscle by CEST MRI at 3T. Magn Reson Med. 2024;91:942-954.
58. Zhou IY, Ji Y, Zhao Y, Viswanathan M, Sun PZ, Zu ZL. Specific and rapid guanidinium CEST imaging using double saturation power and QUASS analysis in a rodent model of global ischemia. Magn Reson Med. 2023.
59. Hilbert T, Nguyen D, Thiran JP, Krueger G, Kober T, Bieri O. True constructive interference in the steady state (trueCISS). Magn Reson Med. 2018;79:1901-1910.
60. Cavallaro M, Coglitore A, Tessitore A, et al. Three-dimensional constructive interference in steady state (3D CISS) imaging and clinical applications in brain pathology. Biomedicines. 2022;10.
61. Liang D, Liu B, Wang JJ, Ying L. Accelerating SENSE using compressed sensing. Magn Reson Med. 2009;62:1574-1584.
62. Breitling J, Korzowski A, Kempa N, et al. Motion correction for three-dimensional chemical exchange saturation transfer imaging without direct water saturation artifacts. NMR Biomed. 2022;35:e4720.
63. Chen H, Chen XR, Lin LJ, et al. Learned spatiotemporal correlation priors for CEST image denoising using incorporated global-spectral convolution neural network. Magn Reson Med. 2023;90:2071-2088.
64. Scheffler K, Lehnhardt S. Principles and applications of balanced SSFP techniques. Eur Radiol. 2003;13:2409-2418.
65. Scheffler K, Hennig J. Is TrueFISP a gradient-echo or a spin-echo sequence? Magn Reson Med. 2003;49:395-397.

Table 1. Signal-to-noise ratios (SNRs) of APTw and NOE maps of egg white phantom acquired using True FISP and FLASH sequences.

| **Sequence** | **APTw map** | **NOE map** |
|---|---|---|
| True FISP | 120.2 | 77.1 |
| FLASH | 71.4 | 49.1 |

Table 2. Contrast-to-noise ratios (CNRs) of APTw and NOE maps of egg white phantom acquired using True FISP and FLASH sequences.

| **Sequence** | **APTw map** | **NOE map** |
|---|---|---|
| True FISP | 42.9 | 56.0 |
| FLASH | 18.3 | 19.8 |

Table 3. CEST values from two sequences in the solid tumor region and the contralateral normal white matter region (nWM) in a patient with a brain tumor.

| **Sequence** | **APTw** | | **NOE** | | **MT** | |
|---|---|---|---|---|---|---|
| | **Solid tumor** | **nWM** | **Solid tumor** | **nWM** | **Solid tumor** | **nWM** |
| True FISP | 0.101±0.030 | 0.068±0.019 | 0.127±0.022 | 0.132±0.024 | 0.241±0.088 | 0.498±0.205 |
| FLASH | 0.104±0.027 | 0.069±0.019 | 0.116±0.019 | 0.126±0.021 | 0.227±0.084 | 0.461±0.178 |

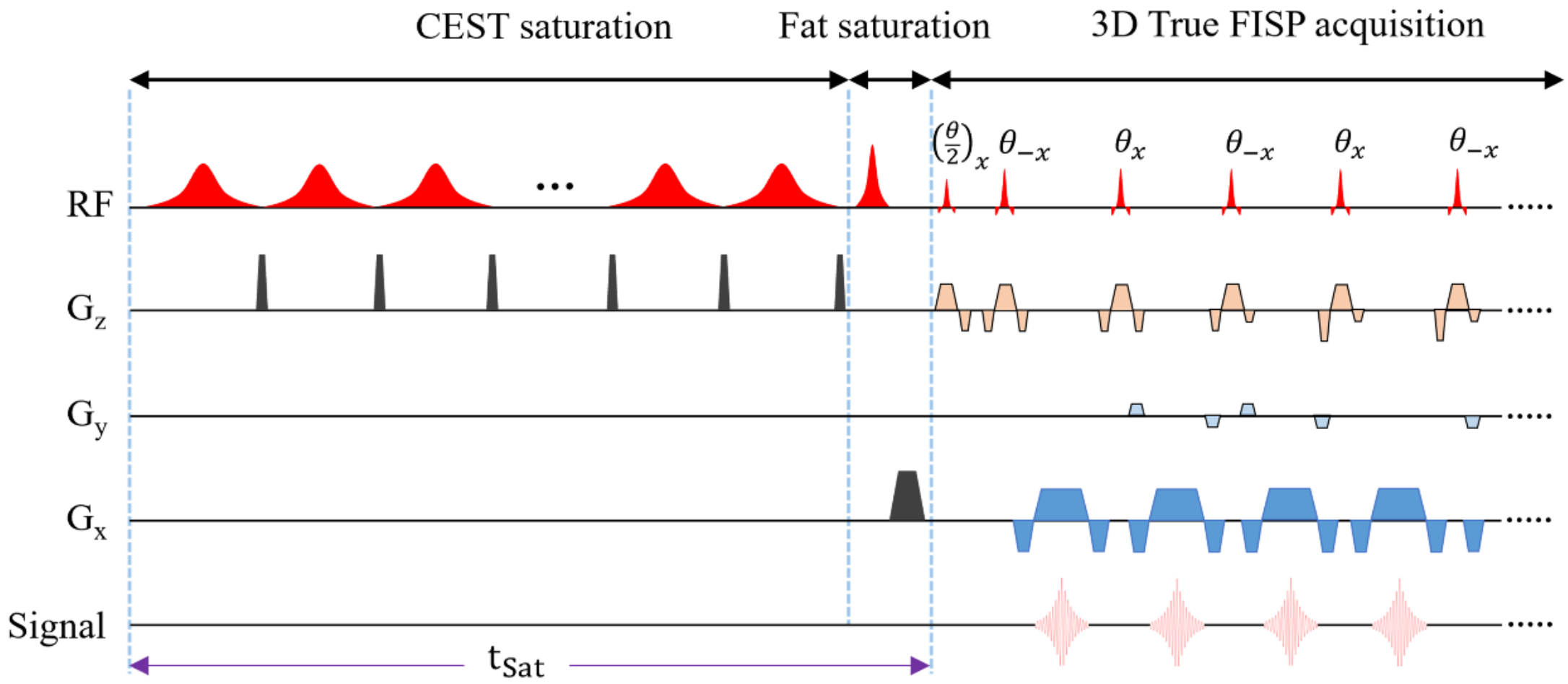

Figure 1. A schematic diagram of the 3D single-shot True FISP CEST sequence. The pulse sequence consists of a train of Gaussian RF pulses for CEST saturation, a fat saturation RF pulse, and a 3D True FISP imaging module. The sign of the flip angle indicates the phase reversal of adjacent RF pulses. The first RF excitation of the True FISP sequence serves as the preparation pulse, with signal acquisition beginning from the second RF excitation. $G_x$, $G_y$, and $G_z$ refer to the frequency-encoding, phase-encoding, and slice-encoding gradients, respectively. RF refers to the radio-frequency pulse.

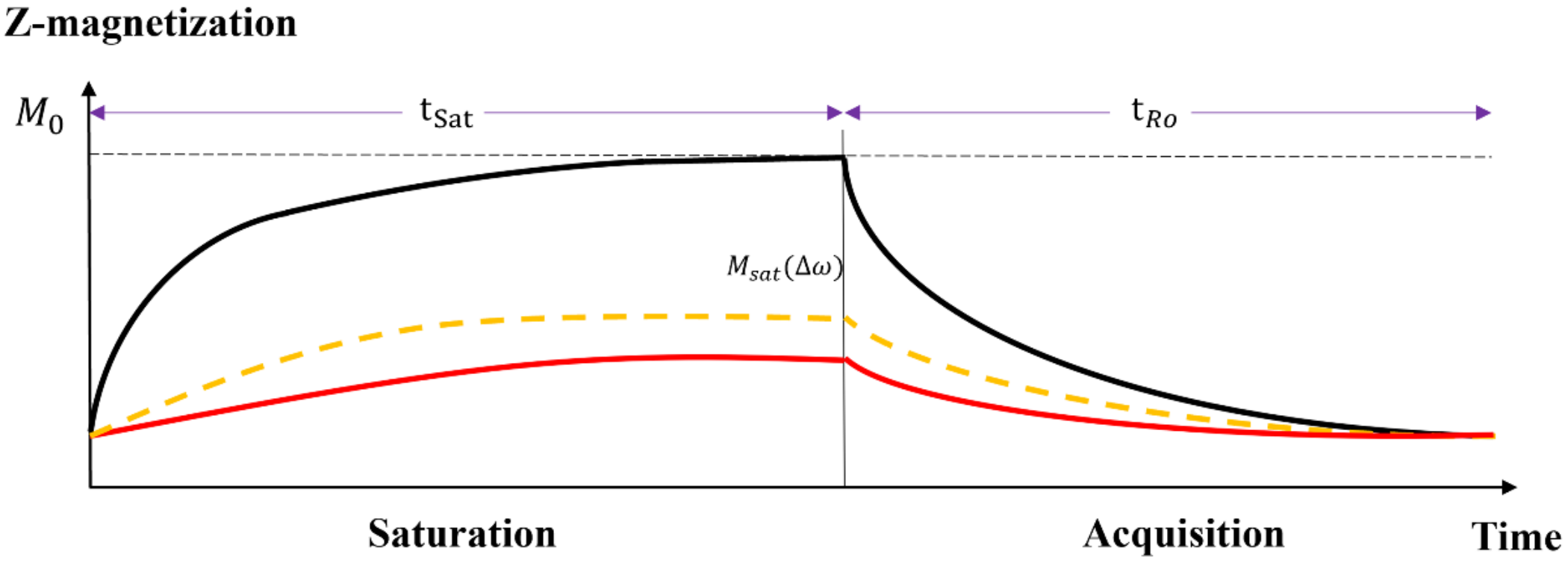

Figure 2. Longitudinal magnetization time curve for single-shot GRE CEST MRI. Following the preparation module, the longitudinal magnetization $M_{sat}$ of the CEST label scan (solid red line) is lower than that of the reference scan (without CEST effect, dashed yellow line). The black solid line illustrates the longitudinal magnetization in the absence of magnetization transfer effects (unsaturated).

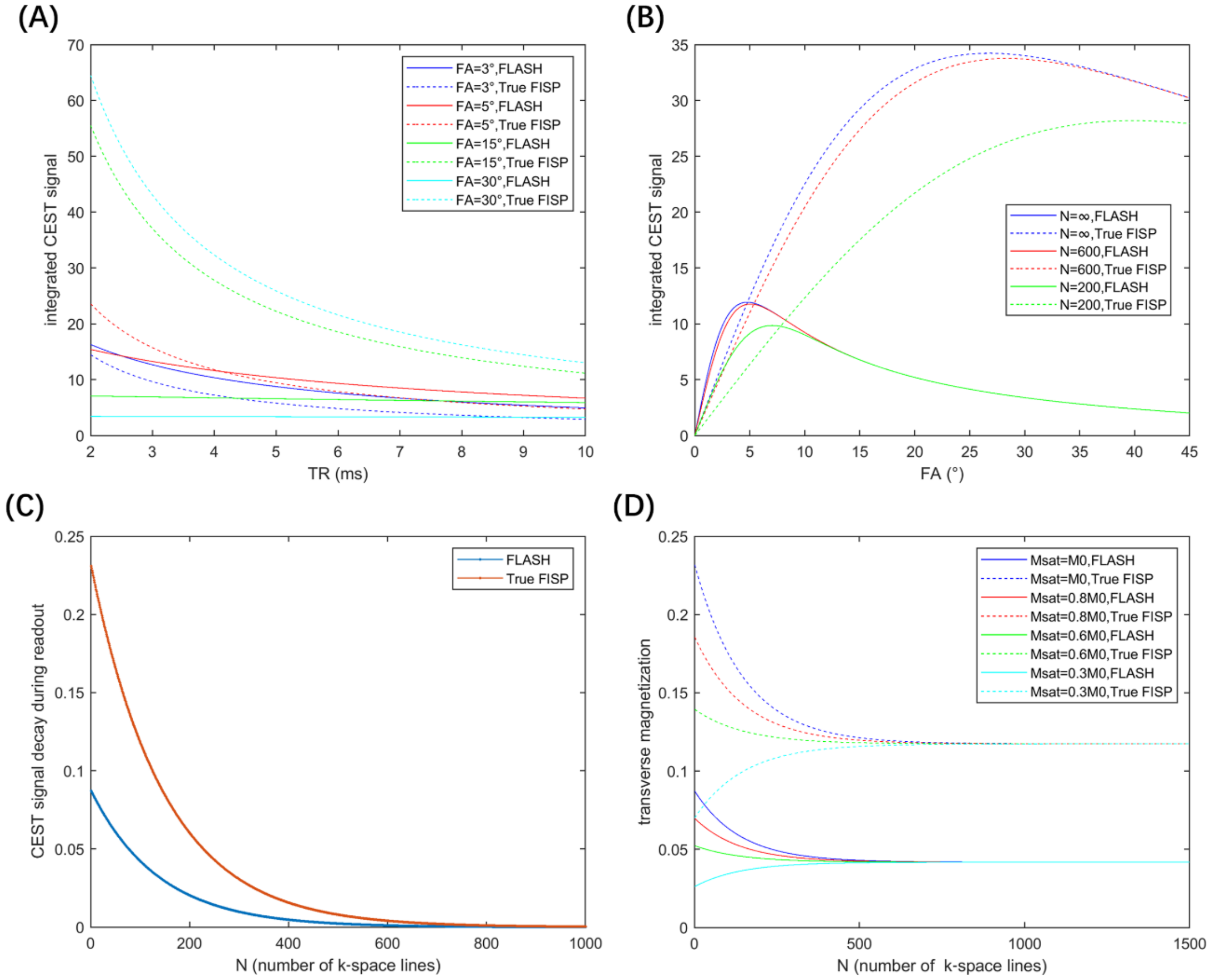

Figure 3. Magnetization simulation analysis in the white matter. (A) The integrated CEST signal (Eq. 3) as a function of the repetition time (TR) for two readout sequences and various flip angles (FAs) under the condition of $N = \infty$. (B) Relationships between the integrated CEST signal and FA for different sequences and numbers of acquired k-space lines. (C) The attenuation curve of the CEST signal (Eq. 2) with acquired K-space lines for different sequences. (D) Evolution of transverse magnetization (Eq. 1) of different sequences under different initial longitudinal magnetization values ($M_{sat}$). $M_{sat} = M_0$ represents the unsaturated condition.

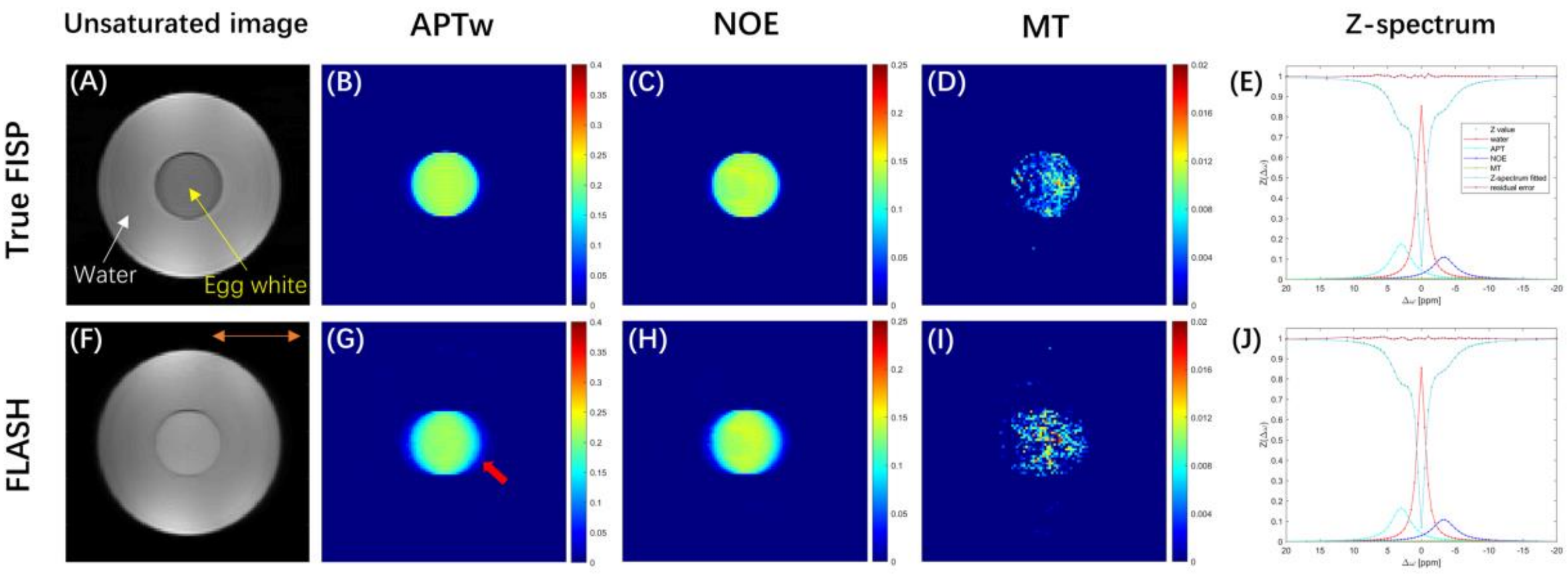

Figure 4. Unsaturated images, APTw, NOE, MT maps and examples of fitted Z-spectrum from the egg white phantom acquired using True FISP (A-E) and FLASH (F-J) sequences. The single-voxel Z-spectrum were randomly selected from voxels at the same position in the first scan of the two sequences. The orange double-headed arrow indicates the direction of phase encoding.

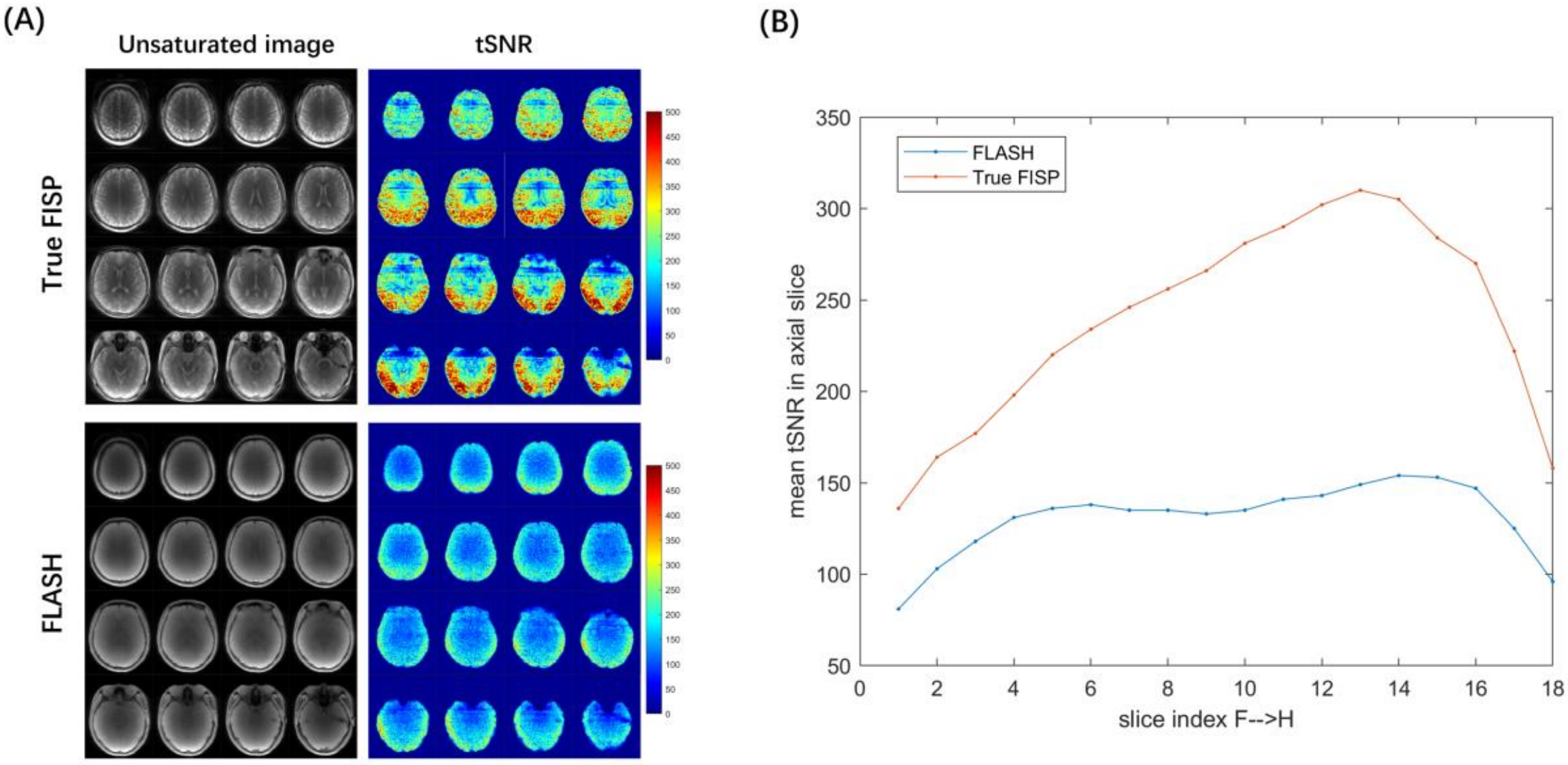

Figure 5. (A) Unsaturated images and temporal signal-to-noise ratios (tSNRs) of saturated images at 3.5 ppm of a healthy subject's brain, acquired using True FISP and FLASH sequences. (B) Mean tSNRs in the axial slices obtained from True FISP and FLASH sequences.

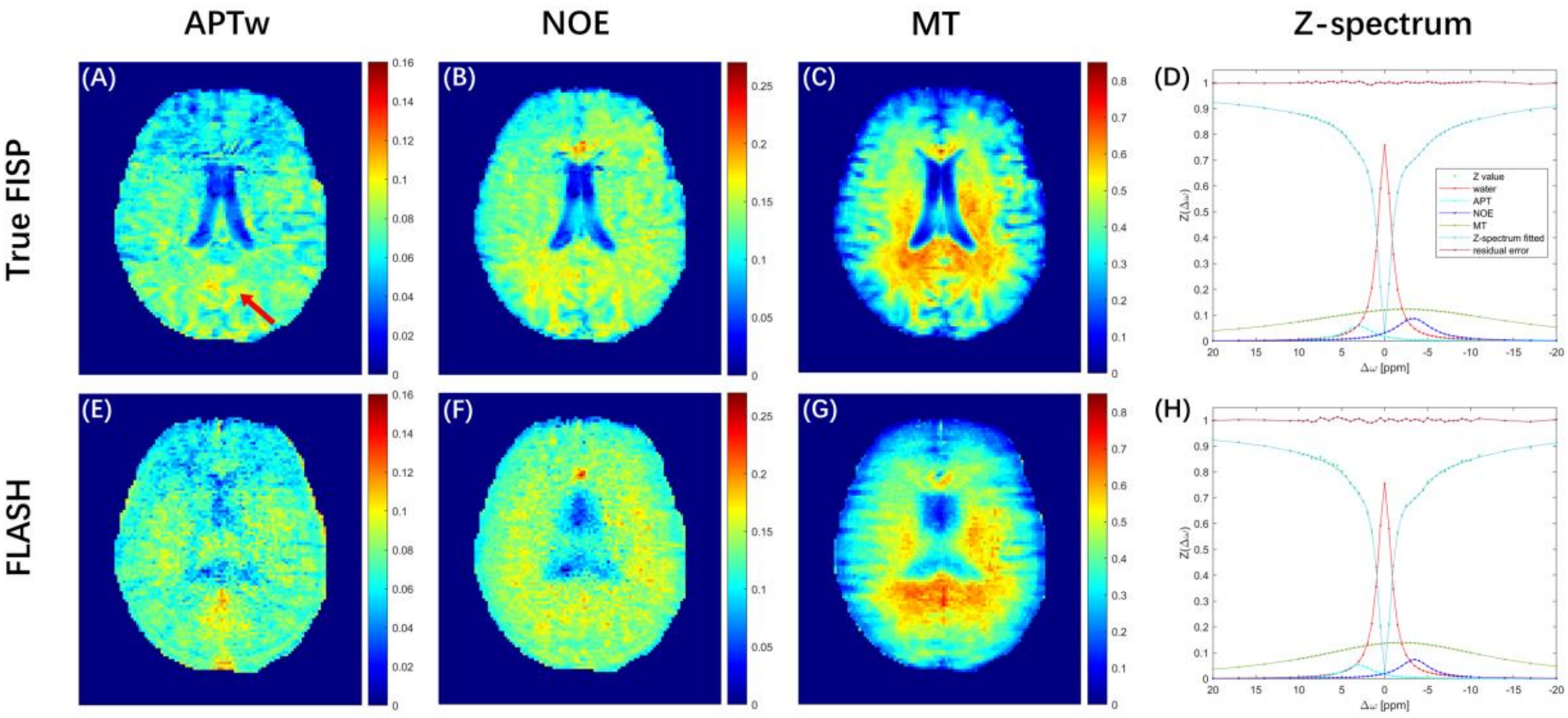

Figure 6. CEST maps and examples of fitted Z-spectrum in the occipital lobe of a healthy subject acquired using the True FISP sequence (A-D) and the FLASH sequence (E-H). The position of single-voxel Z-spectrum was indicated by the red arrow. The location of the voxel to which the Z-spectrum belongs is indicated by the red arrow.

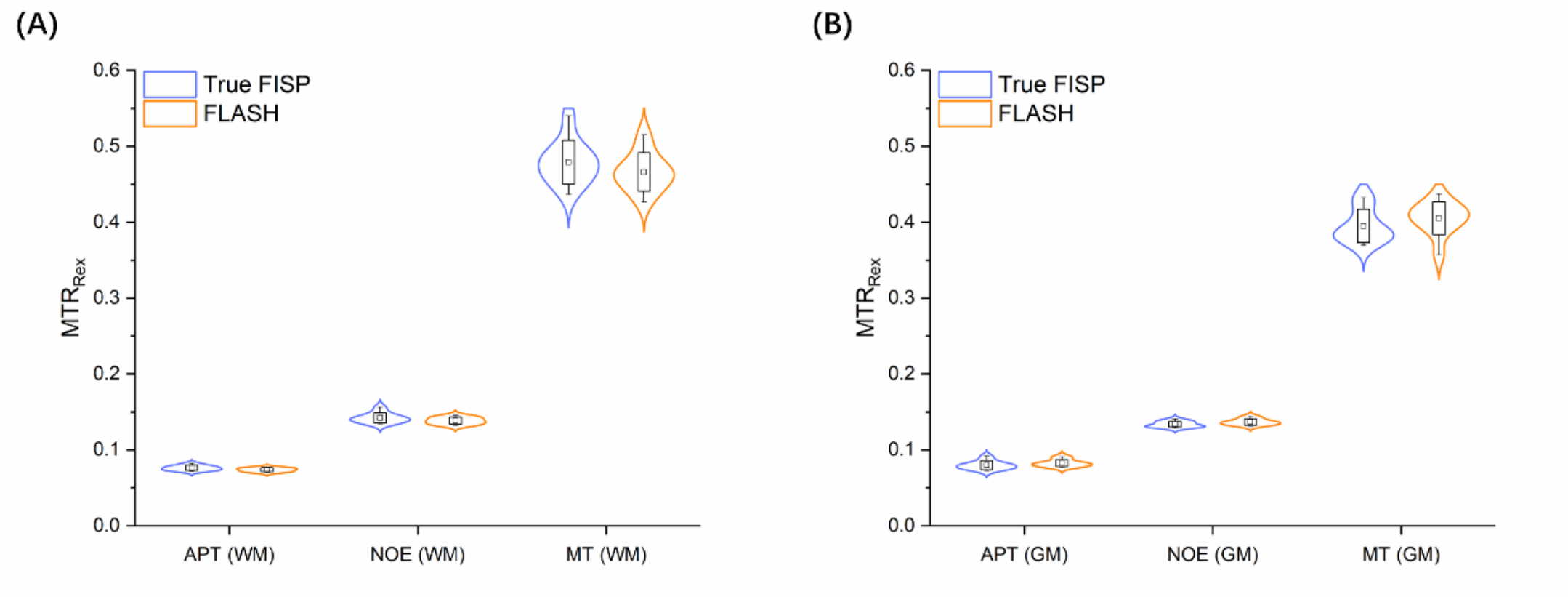

Figure 7. The average $MTR_{Rex}$ values in the white matter (WM, A) and gray matter (GM, B) regions, acquired from ten healthy volunteers using True FISP and FLASH sequences.

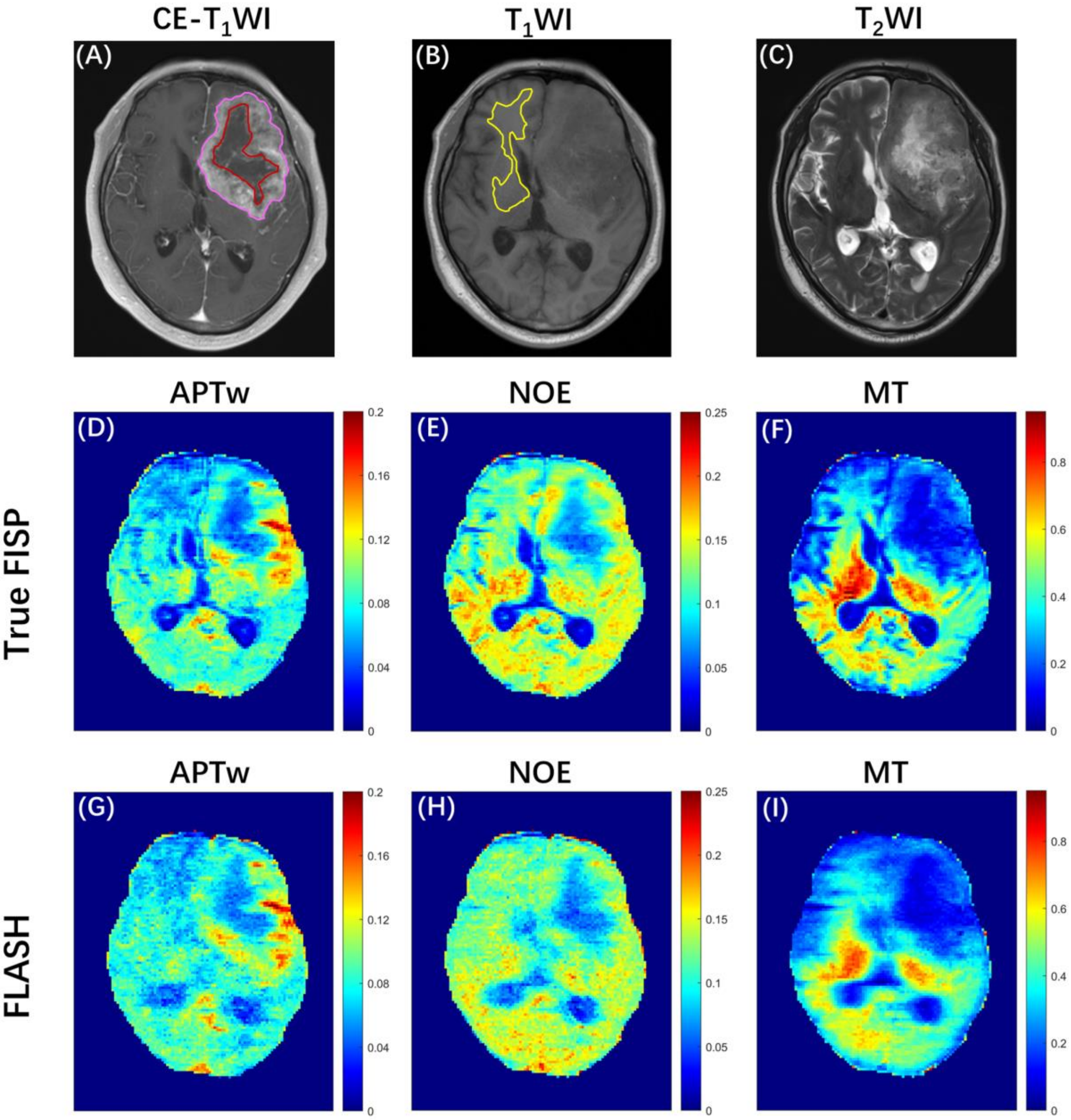

Figure 8. Conventional MRI images (A-C) from a patient with a brain tumor, alongside CEST images from the True FISP sequence (D-F) and the FLASH sequence (G-I). The outer and inner boundaries of the enhanced region are drawn with pink and red lines, respectively (A). The normal appearing white matter in the contralateral hemisphere is outlined in yellow (B).

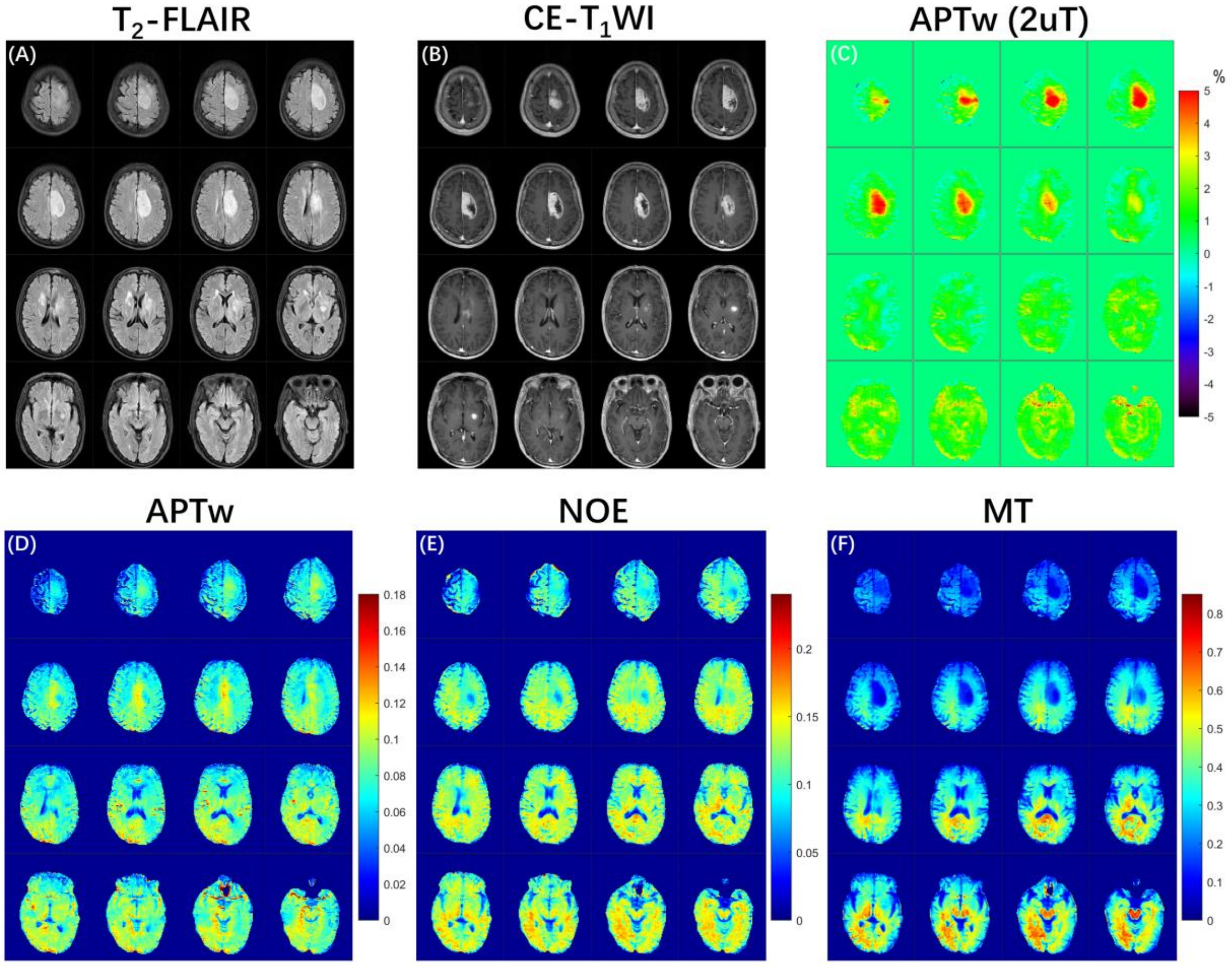

Figure 9. Comparative APTw (2 uT) and multi-pool CEST imaging in a brain tumor patient using True FISP sequence. (A-B) Conventional MRI images. (C) APTw (2 uT) maps acquired at 2 μT saturation power (consensus-recommended protocol), demonstrating elevated signal in tumor regions and near-zero baseline in normal tissue. (D-F) Multi-pool CEST maps acquired at 0.6 μT saturation power, maintaining methodological consistency with the main manuscript.

**SUPPORTING INFORMATION**

Table S1. Initial parameters and boundaries of the center frequency (Δ), amplitude (A), and full width at half maximum (W) of all pools in the Lorentz fitting process.

| **Z-spectrum contributor** | **Parameter** | **Starting value** | **Lower bound** | **Upper bound** |
|---|---|---|---|---|
| Water | Δ (ppm) | 0 | -1 | 1 |
| | A | 0.9 | 0.02 | 1 |
| | W (ppm) | 1.4 | 0.3 | 10 |
| Amide | Δ (ppm) | 3.5 | 3 | 4 |
| | A | 0.025 | 0 | 0.2 |
| | W (ppm) | 0.5 | 0.4 | 3 |
| NOE | Δ (ppm) | -3.5 | -4.5 | -2 |
| | A | 0.02 | 0 | 0.4 |
| | W (ppm) | 3 | 1 | 5 |
| MT | Δ (ppm) | -1 | -4 | -0.5 |
| | A | 0 | 0 | 1 |
| | W (ppm) | 25 | 15 | 100 |

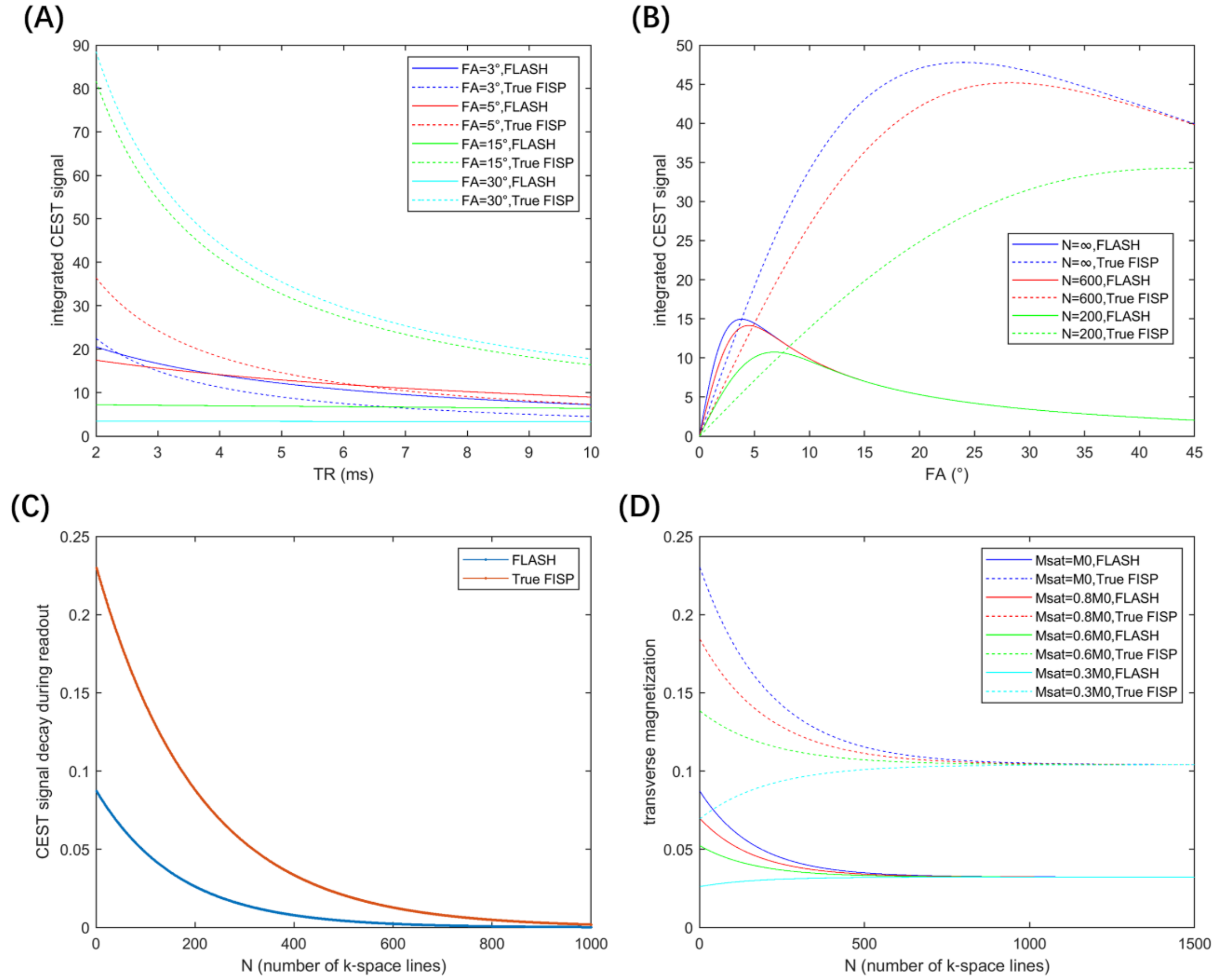

Figure S1. Magnetization simulation analysis in the gray matter. (A) The integrated CEST signal as a function of the repetition time (TR) for two readout sequences and various flip angles (FAs) under the condition of N = ∞. (B) Relationships between the integrated CEST signal and FA for different sequences and numbers of acquired k-space lines. (C) The attenuation curve of CEST signal with acquired K-space lines for different sequences. (D) Evolution of transverse magnetization of different sequences under different initial longitudinal magnetization values ($M_{sat}$). $M_{sat} = M_0$ represents the unsaturated condition.

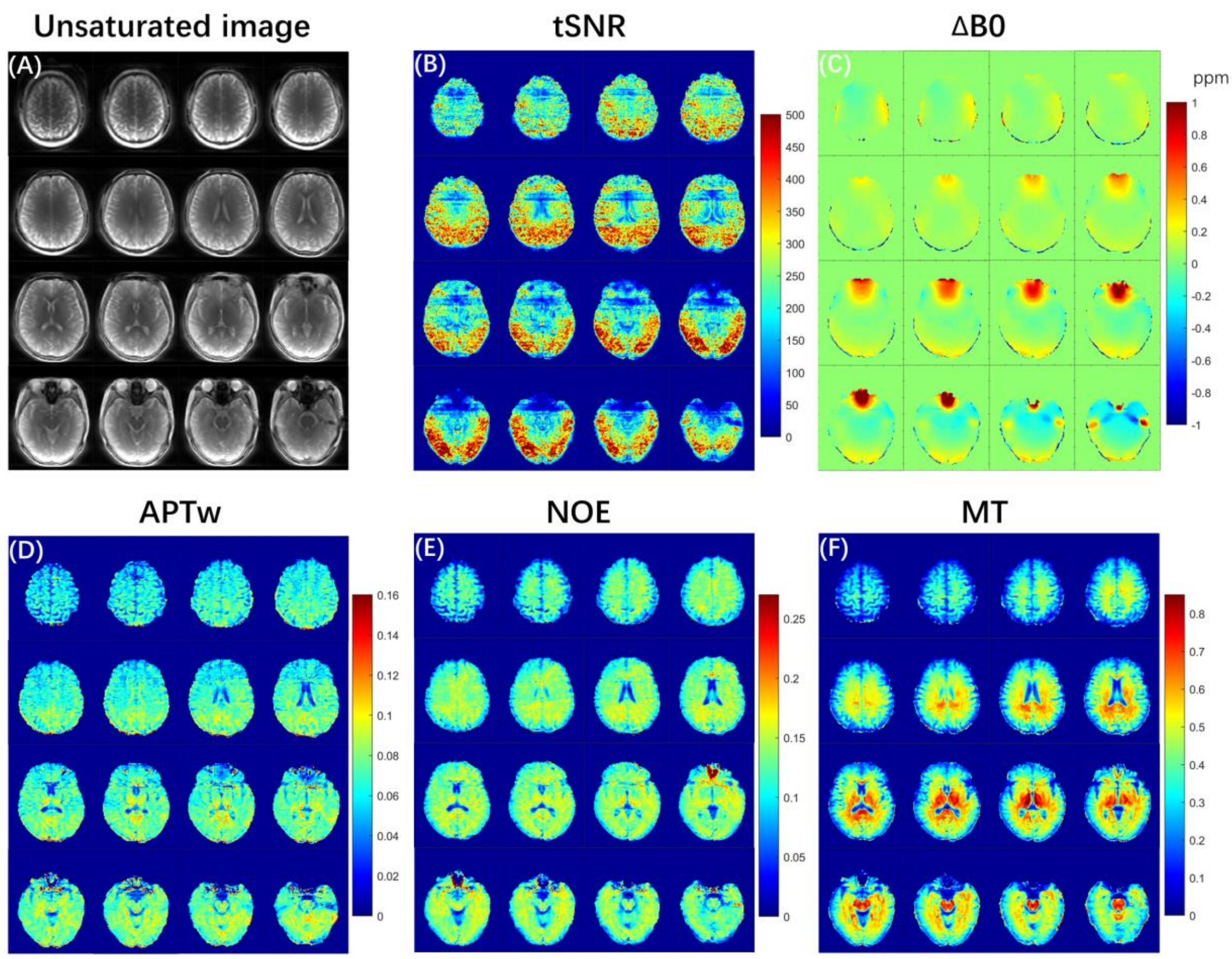

Figure S2. Unsaturated images, tSNR images, ΔB0 maps, along with their corresponding CEST maps of a healthy subject, acquired using the True FISP sequence.

## APPENDIX

### Detailed derivation of the general formula for the evolution of the transverse magnetization of the True FISP sequence

The magnetization response of RF pulses in True FISP sequence can be expressed as:

$$\begin{cases} M_\perp^+ = M_Z^- \cdot \sin\theta - M_\perp^- \cdot \cos\theta \\ M_Z^+ = M_Z^- \cdot \cos\theta + M_\perp^- \cdot \sin\theta \end{cases}. \qquad \text{(A1)}$$

Here, $M_\perp^+$ and $M_\perp^-$ are the transverse magnetization after and before RF excitation,

respectively; $M_Z^+$ and $M_Z^-$ are the longitudinal magnetization after and before RF excitation; and $\theta$ is the flip angle (FA).

Firstly, we assume that the magnetic field within a voxel is extremely uniform, and that the frequency deviation of each voxel from the on-resonance frequency is less than $\pm$1/(2TR). Under these conditions, there is no $T_2$* effect caused by the dephasing of transverse magnetization, and the off-resonance effect is also negligible.

Referring to the description of the magnetization vector evolution of True FISP sequence by Klaus Scheffler et al.,[64] the following mathematical model can be established:

Let $M_{\perp,n}$ represent the transverse magnetization after the nth RF pulse, and $M_{Z,n}$ denote the longitudinal magnetization before the nth RF pulse, with n = 0 indicating the $\theta/2$ preparation pulse.

When n <= 2, insights from the Bloch equation and the pulse mechanism of True FISP sequence lead to the following equations:

$$M_{Z,0} = M_{sat}, \tag{A2}$$

$$M_{Z,1} = M_{Z,0} \cdot \cos(\theta/2) \cdot \sqrt{E_1} + M_0\left(1 - \sqrt{E_1}\right), \tag{A3}$$

$$M_{Z,2} = (M_{Z,1} \cdot \cos\theta + M_{\perp,0} \cdot \sqrt{E_2} \cdot \sin\theta) \cdot E_1 + M_0(1 - E_1), \tag{A4}$$

$$M_{\perp,0} = M_{Z,0} \cdot \sin(\theta/2), \tag{A5}$$

$$M_{\perp,1} = M_{Z,1} \cdot \sin\theta - M_{\perp,0} \cdot \sqrt{E_2} \cdot \cos\theta, \tag{A6}$$

$$M_{\perp,2} = M_{Z,2} \cdot \sin\theta - M_{\perp,1} \cdot E_2 \cdot \cos\theta. \tag{A7}$$

Here, $E_{1,2} = e^{-TR/T_{1,2}}$.

When n >= 3, $M_{Z,n}$ and $M_{\perp,n}$ can be represented as:

$$M_{Z,n} = \left(M_{Z,n-1} \cdot \cos\theta + M_{\perp,n-2} \cdot E_2 \cdot \sin\theta\right) \cdot E_1 + M_0(1 - E_1), \tag{A8}$$

$$M_{\perp,n} = M_{Z,n} \cdot \sin\theta - M_{\perp,n-1} \cdot E_2 \cdot \cos\theta. \tag{A9}$$

Using Eq. (A9), we can derive the following equations:

$$M_{Z,n} = \left(M_{\perp,n} + M_{\perp,n-1} \cdot E_2 \cdot \cos\theta\right)/\sin\theta, \quad \text{(A10)}$$

$$M_{Z,n-1} = \left(M_{\perp,n-1} + M_{\perp,n-2} \cdot E_2 \cdot \cos\theta\right)/\sin\theta. \quad \text{(A11)}$$

Bringing Eqs. (A10) and (A11) into Eq. (A8) can obtain the recursive relationship of the transverse magnetization:

$$M_{\perp,n} = p \cdot M_{\perp,n-1} + q \cdot M_{\perp,n-2} + k, \quad \text{(A12)}$$

where,

$$p = \cos\theta \cdot (E_1 - E_2),$$
$$q = E_1 \cdot E_2,$$
$$k = M_0 \cdot \sin\theta \cdot (1 - E_1).$$

It is easy to see that Eq. (A12) is a second-order non-homogeneous linear recursive sequence, thus, its general term formula can be written:

$$M_{\perp,n} = \left[M_{\perp,1} + \lambda - \frac{M_{\perp,2}+\lambda-\alpha\left(M_{\perp,1}+\lambda\right)}{\beta-\alpha}\right] \cdot \alpha^{n-1} + \frac{M_{\perp,2}+\lambda-\alpha\left(M_{\perp,1}+\lambda\right)}{\beta-\alpha} \cdot \beta^{n-1} - \lambda, \quad \text{(A13)}$$

where,

$$\alpha = \frac{p-\sqrt{p^2+4q}}{2},$$
$$\beta = \frac{p+\sqrt{p^2+4q}}{2},$$
$$\lambda = \frac{k}{p+q-1}.$$

In practical operations, magnetic field inhomogeneity within voxels has a dephasing effect on the transverse magnetization, causing $M_{\perp,n}$ to become a $T_2$*-related value. Interestingly,

Scheffler and Hennig found that under the conditions of $T_1$, $T_2$ > TR and the frequency deviation within the voxel from the on-resonance frequency less than $\pm 1/(2TR)$, the signal of True FISP was refocused at TE = TR/2.[65] This characteristic is similar to the spin echo sequence, where the signal is $T_2$-weighted. Therefore, Eq. (A13) can be used to reflect the signal of True FISP in the case of short TR and relatively small magnetic field offset.